\documentclass[useAMS,usenatbib]{mn2e}
\usepackage[pdftex]{graphicx}
\usepackage{amssymb}
\usepackage{amsmath}
\usepackage{multicol}
\usepackage{subfigure}
\usepackage{hyperref}
\usepackage{lscape}
\usepackage[section]{placeins}

\title[Mass and R/X normalisation correlation]{Influence of mass of black holes on radio/X-ray correlation diagram.}

\author[A. Dusoye]{
Avishek Dusoye \thanks{E-mail:avsdus001@myuct.ac.com}\\}
\begin{document}
\date{}
\pagerange{1-- 9} 
\pubyear{2015}

\maketitle

\label{firstpage}
\begin{abstract}
The radio/X-ray correlation diagram, for Black Hole X-ray Binaries (BHXBs) in the hard state, depicts the connection that might exists between the radio jets and X-ray emitting accretion discs. The current version of the radio/X-ray correlation diagram shows two populations of BHXBs. These two populations evolve along two different correlation tracks ($F_{rad}= \kappa F_{x}^{b}$), namely the standard track and the outliers. Over the past years, the key question has been to explain the existence of these two tracks. In this paper, we investigate the impact of the black hole mass on the radio/X-ray correlation for a sample of 17 BHXBs. We are led to conclude at least one of the following with full consideration of large uncertainities in mass estimates:(i) Most of the reported mass estimates of black holes are incorrect or insufficiently accurate to infer their impact on radio/X-ray correlation diagram. (ii) The estimated radio luminosities and X-ray luminosites are still not reliable enough for two reasons. One reason is due to the lack of associated errors in observational data. Another is that some sources might still transit from one track to the other. (iii) The mass of BH has a significant influence on which track, the source belongs to, on a radio/X-ray correlation diagram. 



\end{abstract}


\section{Introduction}
\hspace{0.7cm} Black hole X-ray binaries (BHXBs) are stellar binary systems consisting of a black hole (BH) and a companion star. They are known to produce X-ray emission through the accretion of mass from the companion star onto the black hole via an accretion disc \citep{remillard2006}, as well as radio emission originating from their jets \citep{Fender2006}. Each component of a BHXB (accretion disc and jets) dominates a specific frequency range, as observed in its spectrum \citep{Markoff2003}. The jets and accretion disc appear pronounced in radio and X-ray respectively. Therefore an approach to investigate the connection between jets and accretion disc for these BHXBs is to study the correlations between radio and X-ray emission, which should be observed quasi-simultaneously. 

In 2003, during the hard state of GX 339-4, \cite{corbel2003} discovered a very strong correlation between radio fluxes at 8.6 GHz using the Australian Telescope Compact Array (ATCA), and X-ray fluxes in the band 3-9 keV using the the Rossi X-ray Timing Explorer (RXTE) and BeppoSAX. This correlation takes the form of a powerlaw,
\begin{equation}
 F_{rad} = \kappa \times F_{x}^{b} \;,
 \end{equation}
 with $b=0.706\pm0.011$ and where $F_{rad}$ is the $8.6$  GHz radio flux density, and $F_{x}$ is the $3-9$ keV X-ray flux density. The validity of this correlation was extended to a larger sample of 10 BHXBs \citep{gallo2003}. The correlation between the radio luminosity and the X-ray luminosity for most of the 10 BHXBs in the hard state seems to be consistent with a power law slope of $b \simeq 0.7 $. This suggested a universal correlation which could imply that most BHXBs might behave similarly in terms of disc-jet coupling.

The work was extended then to larger mass scale using a compiled sample of 100 Active Galactic Nuclei (AGN) with supermassive BHs and 8 galactic BHXBs \citep{Melloni2003}. The parameters considered for the AGN were similar to BHXBs. However, in addition to the radio luminosities $L_R$ ($\rm{erg\,s^{-1}}$) and X-ray luminosities $L_X$ ($\rm{erg\,s^{-1}}$), the mass of the BH ($\rm{M_{\odot}}$) has been taken into account to study the correlation, as it cannot be ignored anymore when comparing stellar mass BHs and supermassive BHs. The results of \cite{Melloni2003} seem to indicate the AGN behave very similarly to BHXBs when it comes to disc-jet coupling. Equation (2) is a general form of Equation (1) when the mass scale is considered, 
\begin{equation}
\log L_{R} = (0.6_{-0.11}^{+0.11})\log L_{X} - (0.78_{-0.11}^{+0.09})\log M_{BH} + 7.33_{-4.07}^{+4.05} \; .
\end{equation}
Since the value of $b \simeq 0.7$ in Equation (1) is within the range of values of the coefficient for the first term in Equation (2), the validity of the universal correlation at a even larger mass scale seems to be maintained. This led to the definition of the fundamental plane of (hard state) black hole activity.  \cite{Plotkin2012} improved on the original work of \cite{Melloni2003} to obtain the refined equation below describing the fundamental plane of BH activity. The latter used the reciprocal Equation (3):
\begin{equation}
\log L_{X} = (1.45_{+0.04}^{-0.04})\log L_{R} - (0.88_{+0.06}^{-0.06})\log M_{BH} - 6.07_{+1.10}^{-1.10}.
\end{equation}

Over the past ten years, many BHXBs have been detected with jets of lower luminosity compared to what is predicted by the previous correlation (Equation 1) for the same X-ray luminosity (accretion-powered). Therefore this suggests that some BHXB may have either a large range of jet power or a large range of X-ray luminosity for similar accretion rates. More and more BHXBs in outburst have been found to be located below the standard universal correlation.  The few examples include XTE 1650-500 \citep{Corbel2004}, IGR J17497-2821 \citep{Rodriguez2007} and Swift J753.5-0127 \citep{Soleri2010}. Given the fact that another population of BHXBs seems to situated off from the standard correlation, these sources are sometimes referred as ``outliers''.

 Statistical evidence has been provided by \cite{Gallo2012} to discard the hypothesis of a single universal correlation between the radio and X-ray luminosity of hard state BHXBs. \cite{Corbel2013} updated the radio/X-ray correlation plot with the new data set of GX 339-4 and some additional sources. The correlation index of $b \sim 0.7$ still remains consistent for the standard correlation on the updated plot. For the outliers, \cite{Coriat2011} found a steeper correlation with index of $b \sim 1.4$, using a large sample of observations for H1743-322. Below a given luminosity, H1743-322 was found to make a transition from the steep lower track to the upper (standard) track of the correlation. Similarly since then, few other ``outliers'' sources have been found to make this transition between the two correlation tracks. These sources are MAXI J1659-152 and XTE J1752-223 (\citeauthor{Jonker2012}, \citeyear{Jonker2012}; \citeauthor{Ratti2012},\citeyear{Ratti2012})
 

\subsection{Definition of problem}

\hspace{0.4cm}The existence of dual correlation tracks completely turns down the original paradigm of the scale-invariant and universal accretion disc jet coupling model. Since this discovery, \textit{a key question is to explain the existence of the two tracks}. One school of thought proposes that black holes with the same accretion inflow, can undergo outflows which can vary over a large range of power. This suggests that some parameters of the BHXBs are actually adjusting the mode of accretion disc-jet coupling. The BH is characterized by three parameters which are its mass, spin and charge. The  charge is considered unimportant since the BH is electrically neutral on average on a macroscopic scale. In addition, there are also parameters in the BHXBs as a binary system such as orbital period, Roche lobe size and inclination which could play a role. Therefore there is a need to constrain those possible parameters in order to provide a plausible explanation for these two tracks.

The radio normalisation (parameter $\kappa$ in Equation 1) serves as an indicator of the position of a source on the radio/X-ray correlation diagram. By investigating the relations between the radio normalisation of a sample of sources with any of the parameters mentioned above, one can study the influence of this parameter on the dual track issue. \cite{Fender2010} show evidence for no correlation between the radio normalisation and BH spin assuming the measurements are correct. \cite{Soleri2011} have investigated the other fixed binary parameters such as orbital period, accretion disc size, and inclination for 17 BHXB systems. In all cases, no evidence for a correlation between these parameters and the radio normalisation has been found. This suggests that they do not play a significant role in the existence of the two correlation tracks.

In this paper, we investigate for plausible explanations for the existence of the dual correlation tracks. This was done by taking the mass estimates for a sample of 17 known BHXBs and attempt to correlate them with radio/X-ray normalisation. Following the work of \cite{Soleri2011} and \cite{Fender2010}, we have adopted the same methodology of using the radio/X-ray normalisation to investigate the mass of the BH (analogous to spin, orbital period, and other binary parameters) as a key parameter. 

There will be two interpretations which can follow, once the results are made. Firstly, for a given accretion rate, jets are fainter due to reduce power or reduce radiative efficiency. \cite{Soleri2011} mentioned that many BHXBs have been found to produce fainter jets at a given accretion-powered luminosities than expected from the earlier correlation, and suggested that BHXBs with similar accretion flows can produce a wide range of jet power due to some parameter or factor (possibly the BH mass). Secondly, for a given accretion rate, the accretion flow is radiatively more efficient. The radiative efficiency of an accretion flow is given by $\epsilon = \frac{L}{\dot{M}c^2}$, where $L$ is the total luminosity and $\dot{M}$ is the accretion rate \citep{Shakura1973}. \cite{Coriat2011} proposed that those hard state BHXBs which are radiating efficiently occupies the outliers' track whereas the BHXBs on the standard track are actually radiating inefficiently. Consequently, it would be interesting to see if the radiative efficiency is influenced by the BH mass.


\vspace{-0.5cm}
\subsection{Mass measurement techniques}
\hspace{0.4cm}Accurate measurements of BH masses are critical to verify whether the position of sources on radio/X-ray correlation diagram has dependence on their BH mass. The measurement of mass for a BH is not trivial and still have large uncertainities associated to it. From the literature, we have compiled mass estimates of a sample of BHXBs, which were obtained from two main measurement techniques:

\hspace{0.4cm}(i) the \textbf{Dynamical Method}; The method is based on Kepler's third law of motion. Since BHXBs are similar to single-lined spectroscopic binaries, only the radial velocity curve of the companion star (extracted from a timing series of spectra) is available. From the radial velocity curve, the orbital period ($P_{orb}$) and the radial velocity semi-amplitude ($K_{c}$) may be extracted and used in the \textit{mass function}, which is defined by Equation (4). This is used to obtain the mass of the BH, $M_{X}$. The mass function is given by,

\begin{equation}
f(M) = \frac{K_{c}^{3}P_{orb}}{2\pi G} = \frac{M_{X}^{3}\sin ^{3} i}{(M_{X} + M_{c})^{2}} =\frac{M_{X}\sin ^{3} i}{(1 + q)^{2}} \:.
\end{equation}

In Equation (4), $M_{c}$ is the mass of the companion star, $q$ is the mass ratio $q= M_{c}/M_{X}$ and G is the Newtonian gravitational constant. This expression assumes a circular orbit at an inclination $i$. If the inclination of the system  is not constrained, the mass function provides a solid lower limit to the mass of the black hole even for a mass of the companion $M_{c} = 0$ and an edge-on geometry \citep{Casares2013}.

\vspace{0.1cm}\hspace{0.4cm}(ii) the \textbf{Spectral-timing correlation scaling technique}; 
The inherent features of the X-ray spectrum  and the rapid variability of X-ray luminosity can be correlated to determine the mass of the black hole and its distance from us \citep{Shaposhnikov2010}. This technique of analysis requires (1) a reference source, whose BH mass and distance is known, and (2) a large scalable number of observations which records how the target and the reference source evolves in its spectra over time as they make spectral transitions. There exists a spectral index associated to each frequency at which quasi periodic oscillations (QPO) occur. By plotting an empirical function to the X-ray variablity data, one can infer both a gradient (index) and a normalisation of the power law function used to fit the data. The index and normalisations are correlated to find any coherence arising from the QPO and these are known as \textit{index$-$normalisation correlation patterns}. The \textit{spectral index$-$QPO frequency} is the frequency of the quasi periodic oscillation at which the spectral index is being taken or considered. The spectral index–QPO frequency and index–normalisation correlation patterns are used to calculate the scaling factors with respect to the reference source. The BH mass can then be extracted since the QPO frequency is assumed to depend on the BH mass only. But this is a highly debated assumption since we still do not understand the exact origin of QPOs.

\vspace{-0.5cm}
\section{Results and analysis}
\hspace{0.7cm}In this section, we present our results on whether the BH mass in a BHXB has an influence on the track it follows on the radio/X-ray correlation diagram. This was done by comparing the radio normalisation of a sample of sources against their corresponding estimated BH mass. Futhermore, we reviewed the dependence of binary parameters (orbital period and inclination) as in \cite{Soleri2011}, for our sample of BHXBs which includes additional and recently monitored sources.

The quasi-simultaneous radio and X-ray observations of the sources used in \cite{Corbel2013} were revisited and updated. The X-ray luminosity is in the 1-10 keV band. Sources having only an upper limit recorded ($\leq 10^{33}\;\rm{erg\,s^{-1}}$ in luminosity) are excluded. Table \ref{table2.2} shows 17 BHXBs, for which the radio/X-ray normalisation values were determined. This was done by minimising our fitting function ($\log \,L_{rad} = \log\,\kappa + \log\,L_{X}^{0.6} + \log\,10^{34} $), which will best fit the evolution of both fluxes for all the observations of each source. The value of index $b = 0.6$ has been fixed. The last term $\log\,10^{34}$ is a normalisation factor for the X-ray luminosity, similar to \cite{Fender2010}. The values $\log \kappa$ are recorded in Table \ref{table2.2}. The best-fit parameters are determined using Chi-square value, which is defined as follows:
\begin{equation}
\chi^{2} = \sum_{i=1}^{N-1} D_{i}^{2} = \sum_{i=1}^{N-1} \dfrac{({y_{i} - f(x_{i})})^{2}}{\sigma_{yi}^{2} + \sigma_{xi}^{2}f^{\prime ^{2}}(x_{i})}
\end{equation}  
The values of $y_i$, $x_i$, $\sigma_{yi}$ and $\sigma_{xi}$ correspond to the radio luminosity, X-ray luminosity, and their corresponding errors respectively for the $i^{th}$ observation of one source. The uncertainties on the best-fit parameters are derived from the diagonal elements of the covariance matrix. All the sources whose data have been used are referenced in Table \ref{table2.2}.


\begin{figure}
\vspace{0.5cm}
\includegraphics[width=0.5 \textwidth]{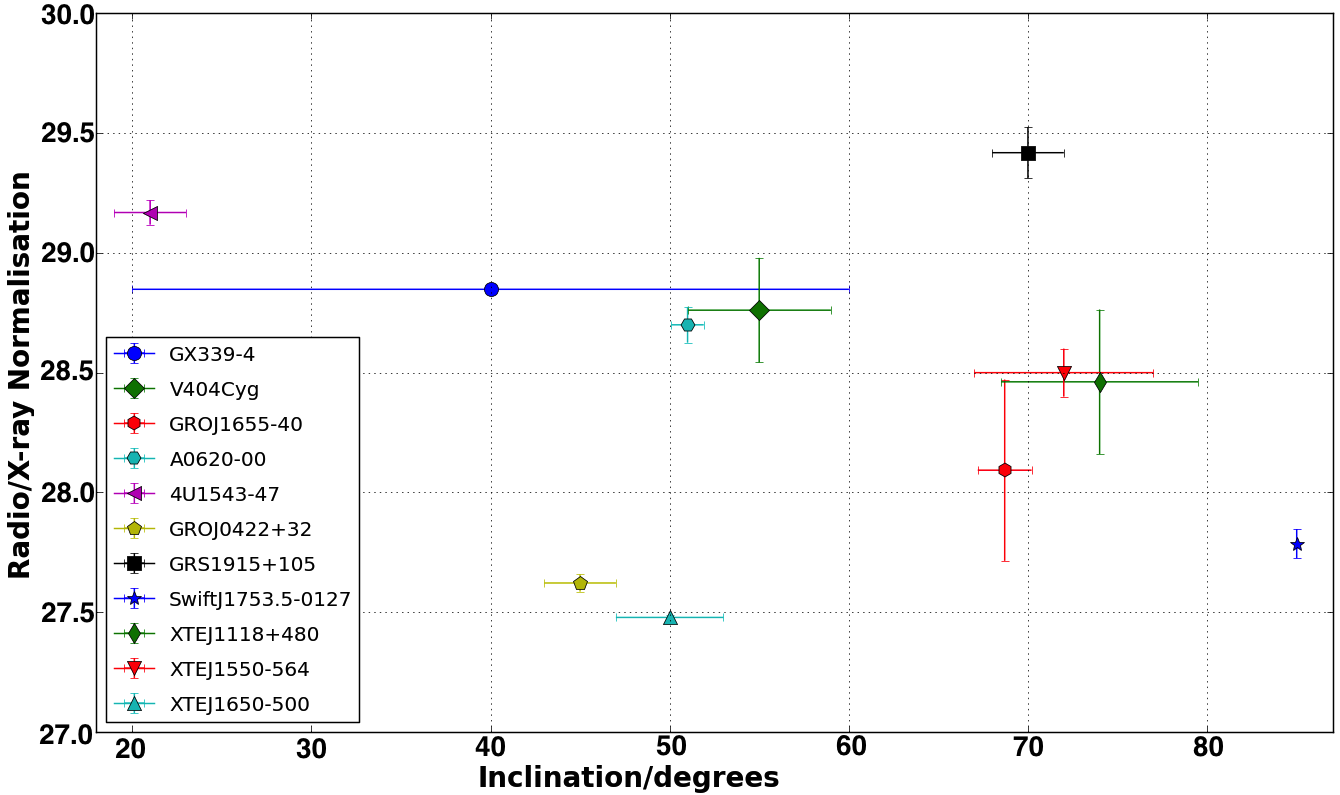}
\includegraphics[width=0.5 \textwidth]{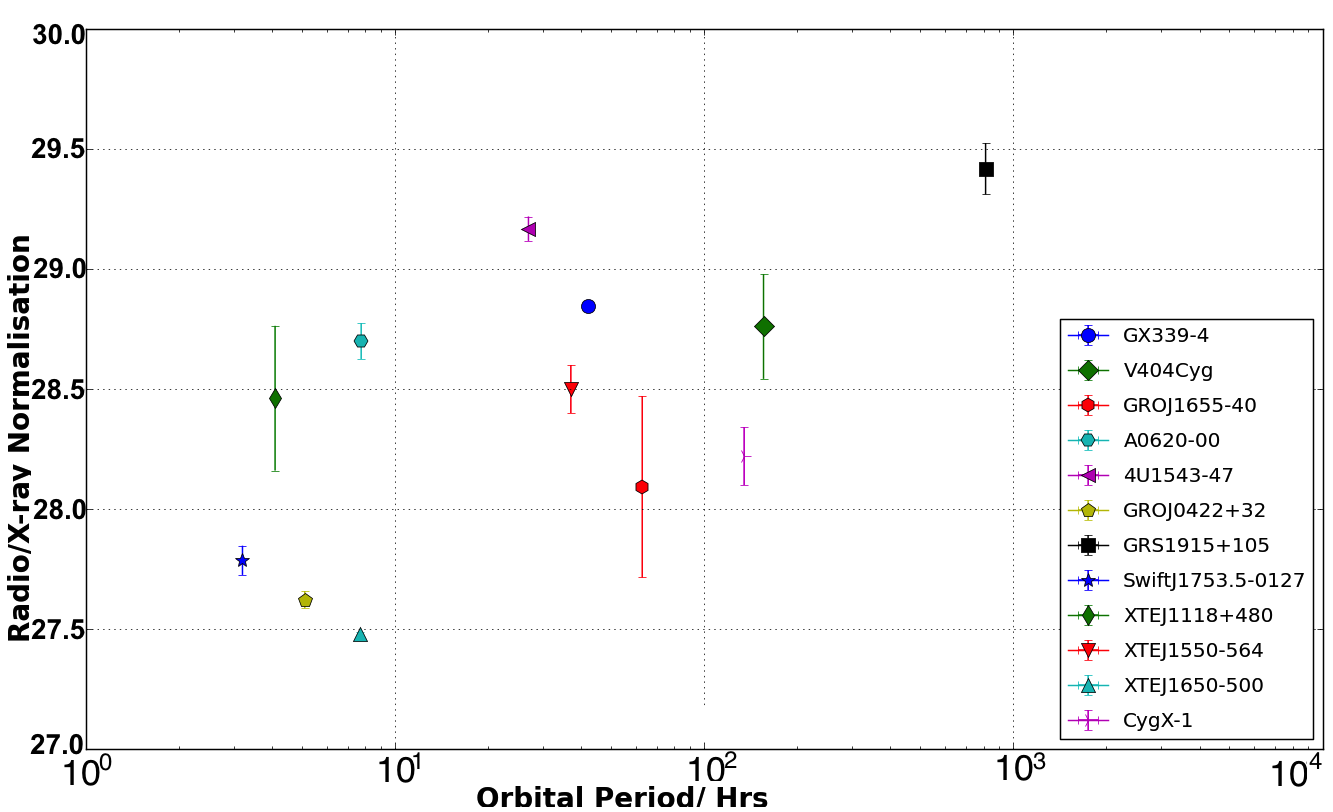}
\caption{Radio/X-ray normalisation values against inclination (top panel) and orbtial period (bottom panel). The Spearman correlation coefficients for these two data sets do not indicate any significant correlation between the parameters (see text).}
\label{A1}
\end{figure}

\begin{figure}
\center
\includegraphics[width=0.5 \textwidth]{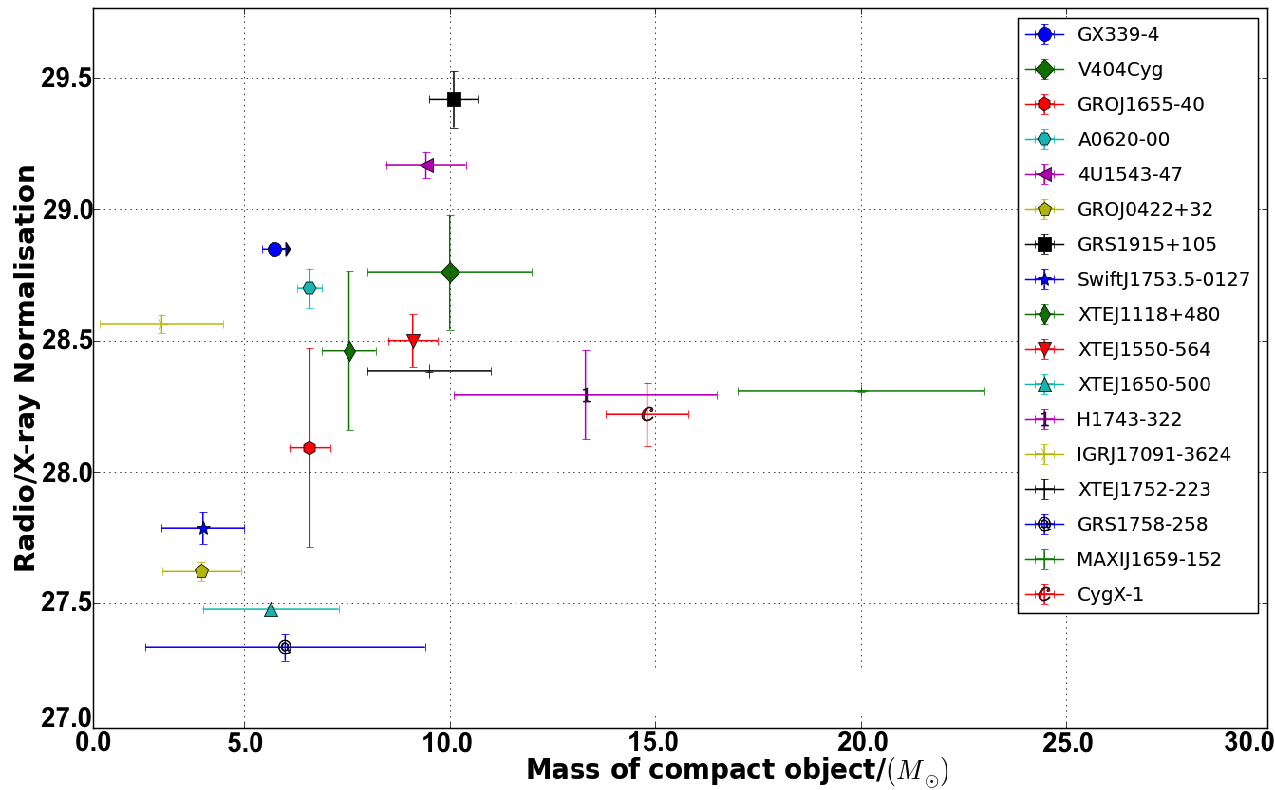}
\caption{Radio/X-ray normalisation as a function of BH mass for the full sample of sources (see Table \ref{table2.2}). The Spearman correlation coefficient is $\Omega \sim 0.28 \pm 0.25$.}
\label{A2}
\end{figure}

\subsection{Inclination and orbital period} 

Using these values of radio normalisation, we can probe which parameter influences the BHXB to follow a particular track. Figure \ref{A1}  shows that there is no clear correlation between the radio/X-ray normalisation with neither the binary inclination nor the orbital period (from literature; see Table \ref{table2.2}). The Spearman correlation coefficient was calculated for the inclination against radio/X-ray normalisation  data set (top panel of Figure \ref{A1}) and found to be $-0.20 \pm 0.55$. Similarly the Spearman correlation coefficient was calculated to be  $-0.27 \pm 0.40$ for the orbital period against radio/X-ray normalisation data set (bottom panel of Figure \ref{A1}). Our results are consistent with the work of \cite{Soleri2011} since the values of the coefficient do not imply a correlation between the studied parameters.

\subsection{Black hole mass estimates - Full sample} 

In order to search for any link between BH mass and radio/X-ray normalisation, we have compiled the most up-to-date mass estimates from the literature for the 17 BHXBs. The values and corresponding references are listed in Table \ref{table2.2}. Figure \ref{A2} shows the radio/X-ray normalisation as a function of the BH mass for all the source of our sample without applying any selection criteria. The Spearman correlation coefficient for the Figure \ref{A2}  is $\Omega \sim 0.28 \pm 0.25$ (without selection criteria). It can noticed that if we include the sources with non-dynamical mass estimates (see  Figure \ref{2.12}), the HMXBs, and sources with quiescent measurement only, the correlation between the normalisation and the mass is much weaker or non-existent. Therefore we require to narrow down this full sample to a selected sample, which is more reliable

\subsection{Black hole mass estimates - Selected sample} 
\hspace{0.5cm} When our sample is restricted using well defined selection criteria, the correlation becomes clearer. The selection criteria which have been adopted are discussed in Section 3.2. Figure \ref{A3} shows our correlation diagram for the radio/X-ray normalisation $\log (\kappa)$ against the BH mass but chosen within the set of full selection criteria mentioned above. The Spearman correlation coefficient was found to be $ \Omega \sim 0.78\pm 0.01$ which means that the set of selection criteria has improved this correlation.

It should be noted that the already known dependence of the radio and X-ray luminosities on the BH mass (i.e. from the fundamental plane of BH activity;\citeauthor{Melloni2003},\citeyear{Melloni2003}) could slightly bias our study. Therefore, we remove this bias by subtracting this mass dependence that originates from the fundamental plane. The \textbf{Corrected normalisation} $\eta$ is simply defined as:
\begin{equation}
 \eta = log(\kappa) + Klog (M/M_{\odot}) \; ,
\end{equation}
 where $K = \frac{\xi_{M}}{\xi_{R}}$ is the ratio of the fundamental plane coefficients as in \cite{Plotkin2012}; $\xi_{M} = -0.88 \pm 0.06$ and $\xi_{R} = 1.45 \pm 0.04$. Having subtracted the mass dependence from the fundamental plane, we can now test for the existence of a correlation between the BH mass and the radio/X-ray normalisation.
 
Figure 5 shows our correlation diagram for the corrected normalisation $\eta$ against the mass of BH for our sample of 11 sources as in Table \ref{table2.2} (within the same selection criteria). The Spearman correlation coefficient was calculated which gives  $ \Omega \sim 0.73 \pm 0.01$. Although by comparing the corrected values and the actual values of radio/X-ray normalisation in Table \ref{table2.2}, one can notice that those corrections, as indicated in Equation 6, are actually small. The correlation still remains when the correction from the fundamental plane of BH activity is taken into account. It should also be noted that the uncertainties of the fundamental plane coefficient (Equation 3) have been propagated into the error of the corrected normalisations. The later are therefore larger than the uncorrected normalisations which reduces the Spearman correlation coefficient. However, the correlation coefficient $\Omega \simeq 0.73 \pm 0.01$ still indicates a significant correlation between the two tested parameters. This result should be further confirmed once additional and more accurate mass measurements will be available. 

\vspace{-0.5cm}
\section{Discussion}
\subsection{Methodology} 
\hspace{0.7cm}Following \cite{Soleri2011} and \cite{Fender2010}, we have adopted the same methodology of using the radio/X-ray normalisation to investigate the influence of the mass of the BH. This methodology seems valid as long as we have (i) accurate BH masses, and (ii) reliable radio/X-ray normalisation values. \\

The BH mass estimates have a cubic dependence on the inclination (Equation 4). Therefore BH mass estimates may not be reliable, given the presence of systematic errors in the measurement of inclination. The two main sources of systematics on light curves are superhump modulation and contamination from rapid aperiodic variability. If there is a large spread of inclination values, there will be more larger range of BH masses \cite{Casares2013}. The BH mass estimates has also two limitations for High Mass X-ray Binaries (HMXBs). Firstly, the value of mass ratio `q' is very small (given the spectral type of their companion star; \citeauthor{Podsiadlowski2003}\citeyear{Podsiadlowski2003}), and therefore the BH mass  would not be very accurate. Secondly, the mass transfer is powered by stellar winds \cite{Ninkov1987}.\\
 
The estimated radio/X-ray normalisations could be debated for its accuracy due lack of errors in observational data especially where only one observation was recorded for one source e.g. A0620-00 and IGRJ17091-3624, IGRJ17177-3656 and XTEJ1720-318. An error corresponding to 10$\%$ of observed fluxes was assigned to these 4 sources which corresponds to the typical (conservative) accuracy obtained with the RXTE satellite. It is also worth discussing the significance for the numerical value of the estimated radio/X-ray normalisation. The exact value of the normalisation is highly dependent on the number of data points (observations) used to calculate it. This is particularly true since we are fitting a power-law with a fixed index $b=0.6$ which does not match the observed slope of the lower track. Given the observations of each source is different, the number of data points associated to an observed source is also different. Hence the estimated radio/X-ray normalisation is not uniformly significant, even over the selected sample. Therefore, the correlation we observe should rather be considered qualitatively rather than quantitatively in the sense that sources evolving on the lower track tend to have less massive BH. Those facts should be taken into consideration before taking any stance regarding this correlation and any inference about it.

\subsection{Selection Criteria of sources:}

\begin{figure}
\center
\includegraphics[width=0.5 \textwidth]{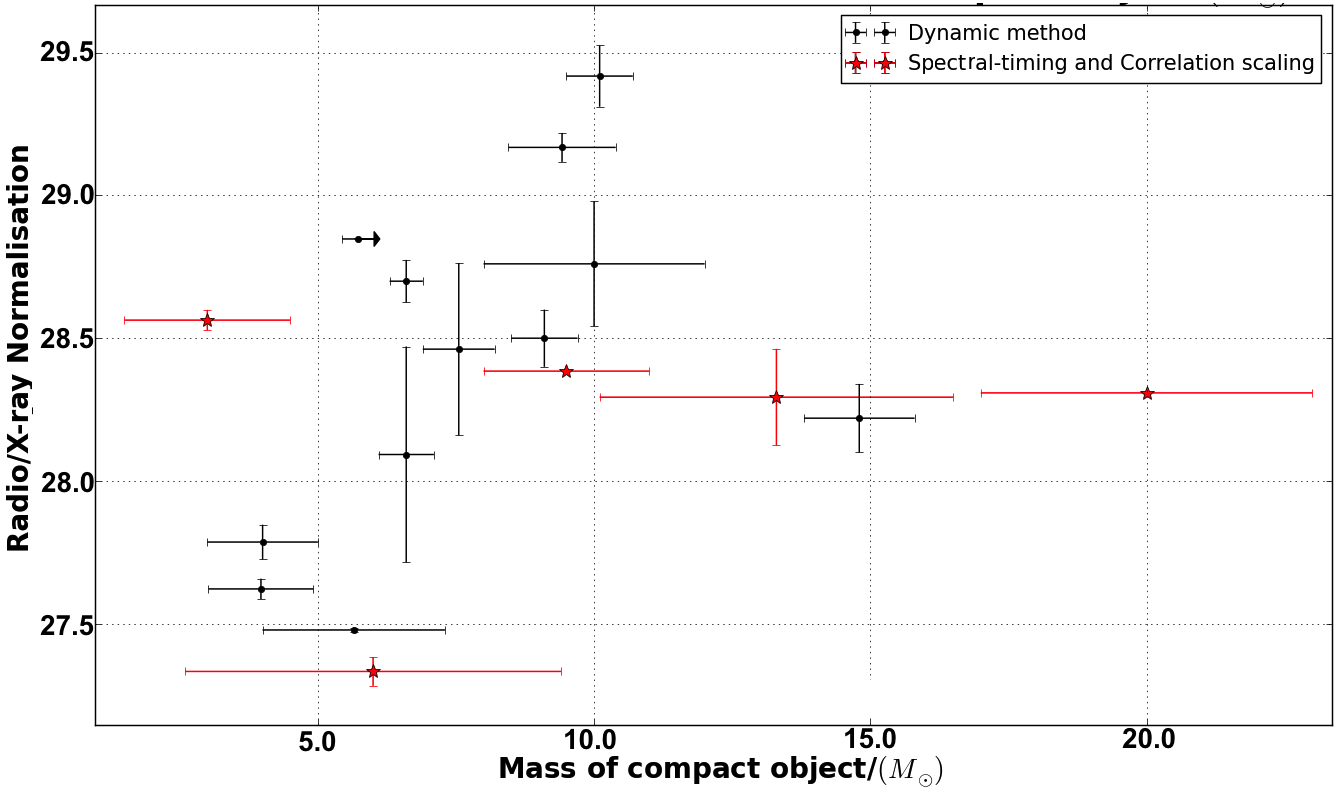}
\caption{Radio/X-ray normalisation as a function of BH mass for the full sample of sources where the measurement technique of the BH mass is specified. These are by (i) Dyanmical method (indicated with black symbols) and (ii) Spectral-timing correlation scaling method (indicated with red pointers).}
\label{2.12}
\end{figure}

\begin{enumerate}
\item \textbf{No sources having just quiescent measurement or single datapoint at low luminosity} \\
Single observation at low luminosity or even quiescent measurements do not show and cannot predict the evolution of the source on a radio/X-ray correlation. It can be biased to associate the sources which are at low luminosity to lie on either the upper or the lower track. The reason is that we have also observed cases  where a source has changed from one track to another e.g. H1743-322. We excluded such sources in our analysis because they cannot be evaluated on which track they lie.
\\
\item \textbf{Low-mass X-ray binaries only} \\
We have excluded Cyg X-1 as it is a HMXB  with a very peculiar behaviour partly due to the fact that it transfers mass through winds. Since Cyg X-1 is also the only HMXB of our sample and it might be biased to include it. 
\\
\item \textbf{Dynamically estimated masses} \\
We additionally excluded sources for which the mass measurement has been obtained through the spectral-timing correlation technique (see Section 1.2). The latter relies on assumptions about a direct link between the BH mass and the spectral and timing properties which are not demonstrated yet. The dynamic method which simply relies on direct observation of lightcurves, is a more reliable and accurate method to obtain the BH masses. Figure \ref{2.12} shows the sources whose mass were estimated dynamically, and the sources whose mass were estimated through spectral-timing correlation.

\end{enumerate}

\subsection{Inference of results}
\hspace{0.7cm} We have found a good correlation for the mass of BH with the radio/X-ray normalisation (see Figure \ref{A3}), where the Spearman rank correlation coefficient is found to be $\Omega = 0.78 \pm 0.01$ ($\Omega = 0.0$ means no correlation and $\Omega = \pm 1.0$ implies an ideal correlation). When we reviewed the correlation of radio/X-ray normalisation with inclination and orbital period, we find the values $ \Omega \simeq -0.20\pm 0.55$ and  $\simeq -0.27 \pm 0.40$ respectively. The results are consistent with the work of \cite{Soleri2011}  since the values of the coefficient do not indicate a significant correlation. The Spearman correlation coefficient is considered as a reliable tool to verify the strength of a correlation. Therefore, this suggests that the mass of BH has a significant influence on which track (standard or outliers) the source belongs to on a radio/X-ray correlation diagram.

Our sample consists of only 17 sources (and 10 sources once the selection criteria have been applied) and may not be fully representative of the general behaviour of all BHXBs that exist. However, we can deduce the following. From the correlation diagram in Figure \ref{A3}, one can infer that sources with more (less) massive BH tend to have higher (lower) radio/X-ray normalisation $\log (\kappa)$. This may imply that BHXBs with a massive BH will tend to populate the standard track whereas BHXBs with a less massive BH will tend to populate the outliers' track. 

We find that, at a given X-ray luminosity, BHXBs with a more massive BH (on the standard track) are observed with higher emission of radio luminosities. Similarly, BHXBs with less massive BH (on the outliers' track) are observed with lower emission of radio luminosities. This may imply that BHXBs with more (less) massive BH will have more (less) powerful jets.

An alternative possibility to the less/more powerful jets (as mentioned above) is that hot inner accretion flow seems to be radiatively more (less) efficient if the BHXB has a less (more) massive accreting BH. This should be valid, at least, within a given range of luminosity (typically above $10^{-3} L_{edd}$), since at low luminosity, sources have been observed to change track (e.g. H1743-322). At the same time, although there exists transitions between the two correlation tracks, further investigations are needed to understand why such a small difference of mass could lead to such noticeable differences in correlation index on the radio/X-ray correlation diagram. A theoretical approach would be necessary to investigate this possibility but that it is out of the scope of this paper.

In addition, \cite{Xie2012} states that the radiative efficiency of accretion flow increases with increasing accretion rate, which gets intensified by viscous interactions. However, since the viscosity parameter of the disc is not related to mass of BH, it could suggest that viscosity plays no critical role in explaining the dual radio/X-ray correlations, if the mass dependence reported here holds.

\vspace{-0.5cm}
\section{Conclusion}
\begin{figure*}
\vbox to 110mm{ \includegraphics[width=0.9 \textwidth]{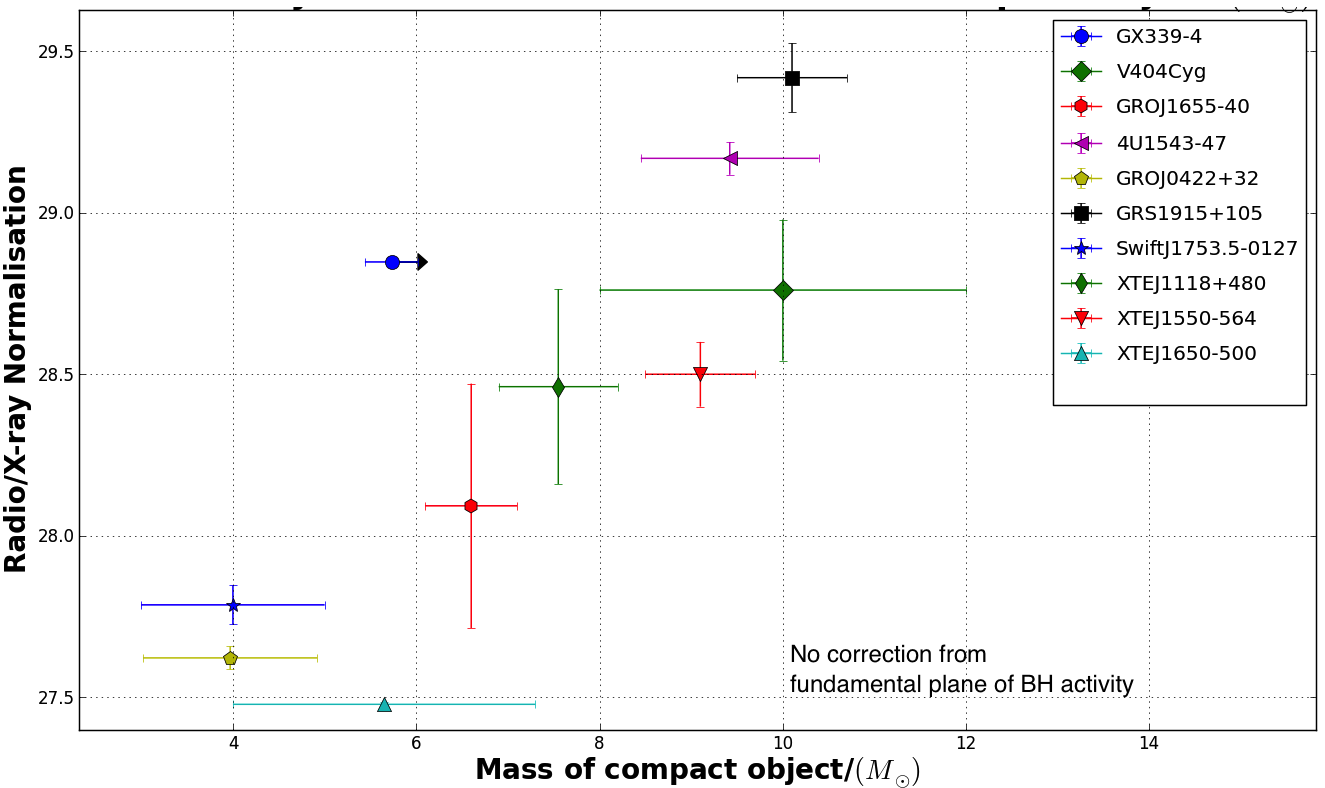}
\caption{Radio/X-ray normalisation against the mass of the accretor for sources listed in Table \ref{table2.2} chosen within selection criteria (see text). The Spearman rank correlation coefficient is calculated as $ \Omega \sim 0.78 \pm 0.01$.}}
\label{A3}
\end{figure*}

\hspace{0.7cm} We investigated three parameters (inclination, orbital period, mass of compact core) with our sample of 17 BHXBs in an attempt to explain the dual tracks on radio/X-ray correlation diagram. Figure \ref{A1} shows that there is no clear correlation of these radio/X-ray normalisation values with inclination and orbital period respectively. Our results are consistent with the work of \cite{Soleri2011}. Figure \ref{A3}  shows all our sources for which we have a radio/X-ray normalisation $\log (\kappa)$ and mass of the compact object. The Spearman correlation coeficient for the Figure \ref{A3}  is $\Omega \sim 0.78\pm 0.01$ (selected sample). We are led to conclude at least one of the following with full consideration of large uncertainities in mass estimates:(i) Most of the reported mass estimates of black holes are incorrect or insufficiently accurate to infer their impact on radio/X-ray correlation diagram. (ii) The estimated radio luminosities and X-ray luminosites are still not reliable enough for two reasons. One reason is due to the lack of associated errors in observational data. Another is that some sources might still transit from one track to the other. (iii) The mass of BH has a significant influence on which track, the source belongs to, on a radio/X-ray correlation diagram. 

\vspace{-0.5cm}
\section*{Acknowledgments}
\hspace{0.7cm}This research has been funded by NRF/SKA Masters Scholarship Grant No. 82750, Reference: SKH1208229663. The author also  acknowledges the financial support from NRF Grants No.120390, Reference: BSFP190416431035, and No.120396, Reference: CSRP190405427545, and No 101775, Reference: SFH150727131568.

\vspace{-0.5cm}
\bibliographystyle{apj}
\bibliography{paper_draft}

\begin{thebibliography}{46}
\expandafter\ifx\csname natexlab\endcsname\relax\def\natexlab#1{#1}\fi

\bibitem[{{Altamirano} {et~al.}(2011){Altamirano}, {Belloni}, {Linares}, {van
  der Klis}, {Wijnands}, {Curran}, {Kalamkar}, {Stiele}, {Motta},
  {Mu{\~n}oz-Darias}, {Casella}, \& {Krimm}}]{Altamirano2011}
{Altamirano}, D., {Belloni}, T., {Linares}, M., {van der Klis}, M., {Wijnands},
  R., {Curran}, P.~A., {Kalamkar}, M., {Stiele}, H., {Motta}, S.,
  {Mu{\~n}oz-Darias}, T., {Casella}, P., \& {Krimm}, H. 2011, ApJ, 742, L17

\bibitem[{Bezayiff(2006)}]{Nate2006}
Bezayiff, N. 2006, PhD thesis, UNIVERSITY OF CALIFORNIA

\bibitem[{{Brocksopp} {et~al.}(2013){Brocksopp}, {Corbel}, {Tzioumis},
  {Broderick}, {Rodriguez}, {Yang}, {Fender}, \& {Paragi}}]{Brocksopp2013}
{Brocksopp}, C., {Corbel}, S., {Tzioumis}, A., {Broderick}, J.~W., {Rodriguez},
  J., {Yang}, J., {Fender}, R.~P., \& {Paragi}, Z. 2013, MNRAS, 432, 931

\bibitem[{{Cadolle Bel} {et~al.}(2007){Cadolle Bel}, {Rib{\'o}}, {Rodriguez},
  {Chaty}, {Corbel}, {Goldwurm}, {Frontera}, {Farinelli}, {D'Avanzo}, {Tarana},
  {Ubertini}, {Laurent}, {Goldoni}, \& {Mirabel}}]{Cadolle2007}
{Cadolle Bel}, M., {Rib{\'o}}, M., {Rodriguez}, J., {Chaty}, S., {Corbel}, S.,
  {Goldwurm}, A., {Frontera}, F., {Farinelli}, R., {D'Avanzo}, P., {Tarana},
  A., {Ubertini}, P., {Laurent}, P., {Goldoni}, P., \& {Mirabel}, I.~F. 2007,
  ApJ, 659, 549

\bibitem[{{Calvelo} {et~al.}(2010){Calvelo}, {Fender}, {Russell}, {Gallo},
  {Corbel}, {Tzioumis}, {Bell}, {Lewis}, \& {Maccarone}}]{Calvelo2010}
{Calvelo}, D.~E., {Fender}, R.~P., {Russell}, D.~M., {Gallo}, E., {Corbel}, S.,
  {Tzioumis}, A.~K., {Bell}, M.~E., {Lewis}, F., \& {Maccarone}, T.~J. 2010,
  MNRAS, 409, 839

\bibitem[{{Cantrell} {et~al.}(2010){Cantrell}, {Bailyn}, {Orosz}, {McClintock},
  {Remillard}, {Froning}, {Neilsen}, {Gelino}, \& {Gou}}]{Cantrell2010}
{Cantrell}, A.~G., {Bailyn}, C.~D., {Orosz}, J.~A., {McClintock}, J.~E.,
  {Remillard}, R.~A., {Froning}, C.~S., {Neilsen}, J., {Gelino}, D.~M., \&
  {Gou}, L. 2010, ApJ, 710, 1127

\bibitem[{{Casares}(1996)}]{Casares1996}
{Casares}, J. 1996, in Astrophysics and Space Science Library, Vol. 208, IAU
  Colloq. 158, Cataclysmic Variables and Related Objects, ed. A.~{Evans} \&
  J.~H. {Wood}, 395

\bibitem[{{Casares} \& {Jonker}(2013)}]{Casares2013}
{Casares}, J. \& {Jonker}, P.~G. 2013, Space Sci. Rev.

\bibitem[{{Corbel} {et~al.}(2013){Corbel}, {Coriat}, {Brocksopp}, {Tzioumis},
  {Fender}, {Tomsick}, {Buxton}, \& {Bailyn}}]{Corbel2013}
{Corbel}, S., {Coriat}, M., {Brocksopp}, C., {Tzioumis}, A.~K., {Fender},
  R.~P., {Tomsick}, J.~A., {Buxton}, M.~M., \& {Bailyn}, C.~D. 2013, MNRAS,
  428, 2500

\bibitem[{{Corbel} {et~al.}(2004){Corbel}, {Fender}, {Tomsick}, {Tzioumis}, \&
  {Tingay}}]{Corbel2004}
{Corbel}, S., {Fender}, R.~P., {Tomsick}, J.~A., {Tzioumis}, A.~K., \&
  {Tingay}, S. 2004, ApJ, 617, 1272

\bibitem[{{Corbel} {et~al.}(2008){Corbel}, {Koerding}, \&
  {Kaaret}}]{Corbel2008}
{Corbel}, S., {Koerding}, E., \& {Kaaret}, P. 2008, MNRAS, 389, 1697

\bibitem[{{Corbel} {et~al.}(2003){Corbel}, {Nowak}, {Fender}, {Tzioumis}, \&
  {Markoff}}]{corbel2003}
{Corbel}, S., {Nowak}, M.~A., {Fender}, R.~P., {Tzioumis}, A.~K., \& {Markoff},
  S. 2003, AA, 400, 1007

\bibitem[{{Coriat} {et~al.}(2011{\natexlab{a}}){Coriat}, {Corbel}, {Prat},
  {Miller-Jones}, {Cseh}, {Tzioumis}, {Brocksopp}, {Rodriguez}, {Fender}, \&
  {Sivakoff}}]{Coriat2011b}
{Coriat}, M., {Corbel}, S., {Prat}, L., {Miller-Jones}, J.~C.~A., {Cseh}, D.,
  {Tzioumis}, A.~K., {Brocksopp}, C., {Rodriguez}, J., {Fender}, R.~P., \&
  {Sivakoff}, G.~R. 2011{\natexlab{a}}, in IAU Symposium, Vol. 275, IAU
  Symposium, ed. G.~E. {Romero}, R.~A. {Sunyaev}, \& T.~{Belloni}, 255--259

\bibitem[{{Coriat} {et~al.}(2011{\natexlab{b}}){Coriat}, {Corbel}, {Prat},
  {Miller-Jones}, {Cseh}, {Tzioumis}, {Brocksopp}, {Rodriguez}, {Fender}, \&
  {Sivakoff}}]{Coriat2011}
{Coriat}, M., {Corbel}, S., {Prat}, L., {Miller-Jones}, J.~C.~A., {Cseh}, D.,
  {Tzioumis}, A.~K., {Brocksopp}, C., {Rodriguez}, J., {Fender}, R.~P., \&
  {Sivakoff}, G.~R. 2011{\natexlab{b}}, MNRAS, 414, 677

\bibitem[{{Fender}(2006)}]{Fender2006}
{Fender}, R. {Jets from X-ray binaries}, ed. W.~H.~G. {Lewin} \& M.~{van der
  Klis}, 381--419

\bibitem[{{Fender} {et~al.}(2010){Fender}, {Gallo}, \& {Russell}}]{Fender2010}
{Fender}, R.~P., {Gallo}, E., \& {Russell}, D. 2010, MNRAS, 406, 1425

\bibitem[{{Gallo} {et~al.}(2006){Gallo}, {Fender}, {Miller-Jones}, {Merloni},
  {Jonker}, {Heinz}, {Maccarone}, \& {van der Klis}}]{Gallo2006}
{Gallo}, E., {Fender}, R.~P., {Miller-Jones}, J.~C.~A., {Merloni}, A.,
  {Jonker}, P.~G., {Heinz}, S., {Maccarone}, T.~J., \& {van der Klis}, M. 2006,
  MNRAS, 370, 1351

\bibitem[{{Gallo} {et~al.}(2003){Gallo}, {Fender}, \& {Pooley}}]{gallo2003}
{Gallo}, E., {Fender}, R.~P., \& {Pooley}, G.~G. 2003, MNRAS, 344, 60

\bibitem[{{Gallo} {et~al.}(2012){Gallo}, {Miller}, \& {Fender}}]{Gallo2012}
{Gallo}, E., {Miller}, B.~P., \& {Fender}, R. 2012, MNRAS, 423, 590

\bibitem[{{Gelino} \& {Harrison}(2003)}]{Gelino2003}
{Gelino}, D.~M. \& {Harrison}, T.~E. 2003, ApJ, 599, 1254

\bibitem[{{Jonker} {et~al.}(2012){Jonker}, {Miller-Jones}, {Homan}, {Tomsick},
  {Fender}, {Kaaret}, {Markoff}, \& {Gallo}}]{Jonker2012}
{Jonker}, P.~G., {Miller-Jones}, J.~C.~A., {Homan}, J., {Tomsick}, J.,
  {Fender}, R.~P., {Kaaret}, P., {Markoff}, S., \& {Gallo}, E. 2012, MNRAS,
  423, 3308

\bibitem[{{Khargharia} {et~al.}(2013){Khargharia}, {Froning}, {Robinson}, \&
  {Gelino}}]{Khargharia2013}
{Khargharia}, J., {Froning}, C.~S., {Robinson}, E.~L., \& {Gelino}, D.~M. 2013,
  AJ, 145, 21

\bibitem[{{Markoff} {et~al.}(2003){Markoff}, {Nowak}, {Corbel}, {Fender}, \&
  {Falcke}}]{Markoff2003}
{Markoff}, S., {Nowak}, M., {Corbel}, S., {Fender}, R., \& {Falcke}, H. 2003,
  A\&A, 397, 645

\bibitem[{{Merloni} {et~al.}(2003){Merloni}, {Heinz}, \& {di
  Matteo}}]{Melloni2003}
{Merloni}, A., {Heinz}, S., \& {di Matteo}, T. 2003, MNRAS, 345, 1057

\bibitem[{{Mu{\~n}oz-Darias} {et~al.}(2008){Mu{\~n}oz-Darias}, {Casares}, \&
  {Mart{\'{\i}}nez-Pais}}]{Munoz2008}
{Mu{\~n}oz-Darias}, T., {Casares}, J., \& {Mart{\'{\i}}nez-Pais}, I.~G. 2008,
  MNRAS, 385, 2205

\bibitem[{{Ninkov} {et~al.}(1987){Ninkov}, {Walker}, \& {Yang}}]{Ninkov1987}
{Ninkov}, Z., {Walker}, G.~A.~H., \& {Yang}, S. 1987, ApJ, 321, 425

\bibitem[{{Orosz} {et~al.}(2011{\natexlab{a}}){Orosz}, {McClintock},
  {Aufdenberg}, {Remillard}, {Reid}, {Narayan}, \& {Gou}}]{Orosz2011b}
{Orosz}, J.~A., {McClintock}, J.~E., {Aufdenberg}, J.~P., {Remillard}, R.~A.,
  {Reid}, M.~J., {Narayan}, R., \& {Gou}, L. 2011{\natexlab{a}}, ApJ, 742, 84

\bibitem[{{Orosz} {et~al.}(2011{\natexlab{b}}){Orosz}, {Steiner}, {McClintock},
  {Torres}, {Remillard}, {Bailyn}, \& {Miller}}]{Orosz2011}
{Orosz}, J.~A., {Steiner}, J.~F., {McClintock}, J.~E., {Torres}, M.~A.~P.,
  {Remillard}, R.~A., {Bailyn}, C.~D., \& {Miller}, J.~M. 2011{\natexlab{b}},
  ApJ, 730, 75

\bibitem[{{Plotkin} {et~al.}(2012){Plotkin}, {Markoff}, {Kelly}, {K{\"o}rding},
  \& {Anderson}}]{Plotkin2012}
{Plotkin}, R.~M., {Markoff}, S., {Kelly}, B.~C., {K{\"o}rding}, E., \&
  {Anderson}, S.~F. 2012, MNRAS, 419, 267

\bibitem[{{Podsiadlowski} {et~al.}(2003){Podsiadlowski}, {Rappaport}, \&
  {Han}}]{Podsiadlowski2003}
{Podsiadlowski}, P., {Rappaport}, S., \& {Han}, Z. 2003, MNRAS, 341, 385

\bibitem[{{Ratti} {et~al.}(2012){Ratti}, {Jonker}, {Miller-Jones}, {Torres},
  {Homan}, {Markoff}, {Tomsick}, {Kaaret}, {Wijnands}, {Gallo}, {{\"O}zel},
  {Steeghs}, \& {Fender}}]{Ratti2012}
{Ratti}, E.~M., {Jonker}, P.~G., {Miller-Jones}, J.~C.~A., {Torres}, M.~A.~P.,
  {Homan}, J., {Markoff}, S., {Tomsick}, J.~A., {Kaaret}, P., {Wijnands}, R.,
  {Gallo}, E., {{\"O}zel}, F., {Steeghs}, D.~T.~H., \& {Fender}, R.~P. 2012,
  MNRAS, 423, 2656

\bibitem[{{Rebusco} {et~al.}(2012){Rebusco}, {Moskalik}, {Klu{\'z}niak}, \&
  {Abramowicz}}]{Rebusco2012}
{Rebusco}, P., {Moskalik}, P., {Klu{\'z}niak}, W., \& {Abramowicz}, M.~A. 2012,
  AA, 540, L4

\bibitem[{{Remillard} \& {McClintock}(2006)}]{remillard2006}
{Remillard}, R.~A. \& {McClintock}, J.~E. 2006, Annu. Rev. AA, 44, 49

\bibitem[{{Rodriguez} {et~al.}(2007){Rodriguez}, {Cadolle Bel}, {Tomsick},
  {Corbel}, {Brocksopp}, {Paizis}, {Shaw}, \& {Bodaghee}}]{Rodriguez2007}
{Rodriguez}, J., {Cadolle Bel}, M., {Tomsick}, J.~A., {Corbel}, S.,
  {Brocksopp}, C., {Paizis}, A., {Shaw}, S.~E., \& {Bodaghee}, A. 2007, ApJ,
  655, L97

\bibitem[{{Rodriguez} {et~al.}(2011){Rodriguez}, {Corbel}, {Caballero},
  {Tomsick}, {Tzioumis}, {Paizis}, {Cadolle Bel}, \&
  {Kuulkers}}]{Rodriguez2011}
{Rodriguez}, J., {Corbel}, S., {Caballero}, I., {Tomsick}, J.~A., {Tzioumis},
  T., {Paizis}, A., {Cadolle Bel}, M., \& {Kuulkers}, E. 2011, AA, 533, L4

\bibitem[{{Rushton} {et~al.}(2010){Rushton}, {Spencer}, {Fender}, \&
  {Pooley}}]{Rushton2010}
{Rushton}, A., {Spencer}, R., {Fender}, R., \& {Pooley}, G. 2010, AA, 524, A29

\bibitem[{{Shahbaz}(2003)}]{Shahbaz2003}
{Shahbaz}, T. 2003, MNRAS, 339, 1031

\bibitem[{{Shakura} \& {Sunyaev}(1973)}]{Shakura1973}
{Shakura}, N.~I. \& {Sunyaev}, R.~A. 1973, A\&A, 24, 337

\bibitem[{{Shaposhnikov} {et~al.}(2010){Shaposhnikov}, {Markwardt}, {Swank}, \&
  {Krimm}}]{Shaposhnikov2010}
{Shaposhnikov}, N., {Markwardt}, C., {Swank}, J., \& {Krimm}, H. 2010, ApJ,
  723, 1817

\bibitem[{{Shaposhnikov} {et~al.}(2011){Shaposhnikov}, {Swank}, {Markwardt}, \&
  {Krimm}}]{Shaposhnikov2011}
{Shaposhnikov}, N., {Swank}, J.~H., {Markwardt}, C., \& {Krimm}, H. 2011, ArXiv

\bibitem[{{Shaposhnikov} \& {Titarchuk}(2009)}]{Shaposhnikov2009}
{Shaposhnikov}, N. \& {Titarchuk}, L. 2009, ApJ, 699, 453

\bibitem[{{Soleri} \& {Fender}(2011)}]{Soleri2011}
{Soleri}, P. \& {Fender}, R. 2011, MNRAS, 413, 2269

\bibitem[{{Soleri} {et~al.}(2010){Soleri}, {Fender}, {Tudose}, {Maitra},
  {Bell}, {Linares}, {Altamirano}, {Wijnands}, {Belloni}, {Casella},
  {Miller-Jones}, {Muxlow}, {Klein-Wolt}, {Garrett}, \& {van der
  Klis}}]{Soleri2010}
{Soleri}, P., {Fender}, R., {Tudose}, V., {Maitra}, D., {Bell}, M., {Linares},
  M., {Altamirano}, D., {Wijnands}, R., {Belloni}, T., {Casella}, P.,
  {Miller-Jones}, J.~C.~A., {Muxlow}, T., {Klein-Wolt}, M., {Garrett}, M., \&
  {van der Klis}, M. 2010, MNRAS, 406, 1471

\bibitem[{{Steeghs} {et~al.}(2013){Steeghs}, {McClintock}, {Parsons}, {Reid},
  {Littlefair}, \& {Dhillon}}]{Steeghs2013}
{Steeghs}, D., {McClintock}, J.~E., {Parsons}, S.~G., {Reid}, M.~J.,
  {Littlefair}, S., \& {Dhillon}, V.~S. 2013, ApJ, 768, 185

\bibitem[{{Wagner} {et~al.}(1992){Wagner}, {Kreidl}, {Howell}, \&
  {Starrfield}}]{Wagner1992}
{Wagner}, R.~M., {Kreidl}, T.~J., {Howell}, S.~B., \& {Starrfield}, S.~G. 1992,
  ApJ, 401, L97

\bibitem[{{Xie} \& {Yuan}(2012)}]{Xie2012}
{Xie}, F.-G. \& {Yuan}, F. 2012, MNRAS, 427, 1580

\end{thebibliography}

\newpage

\begin{figure*}
\vbox to 110mm{ \includegraphics[width=0.9 \textwidth]{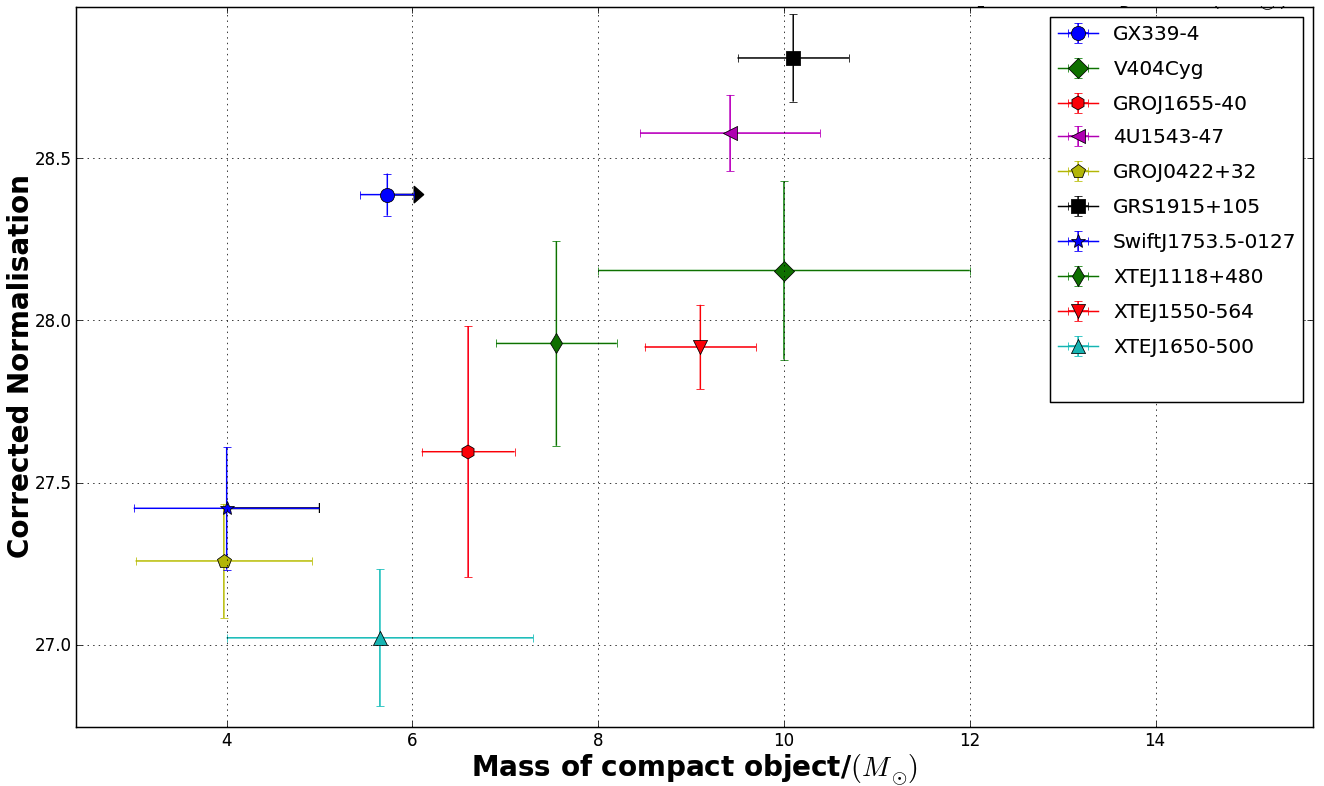}
\caption{The correlation diagram for the corrected normalisation $\eta$ against the mass of the accretor for sources listed in Table \ref{table2.2} chosen within selection criteria. These radio/X-ray normalisations are corrected from any mass dependence from fundamental plane of BH activity (see Equation 6). The Spearman rank correlation coefficient for the corrected version is found to be $\Omega \sim 0.73 \pm 0.01$.}}
\label{A7}
\end{figure*}

\newpage
\begin{landscape}
\thispagestyle{empty} 
\begin{table*}
\centering

\vspace{0.5cm}
\caption{Parameters for the sample of black hole X-ray binaries used in this study; The sources are identified by their names. $\eta$ and $\log \kappa$, represent the corrected and uncorrected radio/X-ray normalisation. $Err_{\eta}$ and $Err_{\log\kappa}$ are the associated errors. $M_{X} \pm dM_{X}$ denotes the mass of the source X. $i \pm di$ denotes inclination and $D \pm dD$ represents the distance to the source. The references for mass estimates as well as radio and X-ray luminosity are under RefA and RefB columns respectively. }
\begin{tabular}{|l|c|c|c|c|c|c|c|c|c|c|l|c|c|l|}
\hline
\bf{Sources} & \bf{$\eta$} & \bf{$Err_{\eta}$} &\bf{$\log\kappa$} &\bf{$Err_{\log\kappa}$}  & \bf{P$_{orb}$/hrs}  & \bf{$M_{X}/M_{\odot}$} & \bf{$dM_{X}$} & \bf{i$/ \circ$} & \bf{di} & \bf{D/kpc} & \bf{dD} & \bf{RefA}  &\bf{RefB}\\
\hline
GX339-4 		 	&28.39	  	&0.07		&28.848 	 	 &0.002		&42.14		&$>$5.73  &0.29	&40	  &20	&8	      &0		&[1]	    	&(a)(b)\\
\hline
V404Cyg 		 	&28.15		&0.274  		&28.76 	 	 &0.22 		&155.31		&10.00	  &2.0	    &55    &4	&2.39	  &0.14	& [2][3]		  	&(d)\\
\hline
GROJ1655-40 	 	&27.59		&0.39		&28.09 	 	 &0.38		&62.92		&6.60	  &0.5 	&68.7  &1.5	 &$<$1.7	   &-	& [4]		&(f)(g)\\
\hline
IGRJ17091-3624	&28.27		&0.33		&28.56	 	 &0.04		&-			&3.00	  &1.5	&-	   &-	 &20.0	   &-	& [5][6]		&(h)\\
\hline
A0620-00		 	&28.20 		&0.10		&28.70	 	 &0.07		&7.75		&6.60	  &0.3	&51	   &0.9	 &1.2	   &0.4	& [7]			&(i)\\  
\hline
4U1543-47		&28.58		&0.12	 	&29.17 	 	 &0.05		&26.95	    &9.42	  &0.97 	&21	   &2	 &7.5	   &0.5	& [8][20]				&(j)\\	 
\hline
GROJ0422+32		&27.26		&0.18		&27.62 	 	 &0.04		&5.09     	&3.97	  &0.95	&45	   &2	 &2.49	   &0.3	& [9]		&(j)\\	
\hline
H1743-322		&27.61		&0.26		&28.29 	 	 &0.17		&-			&13.30	  &3.2	&70	   &-	 &9.1	   &1.5 &  [10]	    &(c)\\
\hline 
GRS1915+105		&28.81		&0.13	   	&29.42	 	 &0.11		&812.40	    &10.10	  &0.6	&70	   &2	 &11.2	   &0.8	& [11]			  &(e)\\
\hline	 
SwiftJ1753.5-0127  &27.42	&0.19		&27.77 	 	 &0.06		&3.20		&4.0 	  &1.0	&85	   &0	 &8	       &-	&[18]		&(k)(l)(m)\\	 	
\hline
XTEJ1118+480		&27.93		&0.32		&28.46	 	 &0.30		&4.08		&7.55	  &0.65	&74	   &5.5	 &1.71	   &0.05	 & [12]		&(j)(n)\\ 
\hline	
XTEJ1550-564		&27.92		&0.13     	&28.50		 &0.10		&36.96 		&9.10	  &0.6	&72    &5	 &4.1      &0.8	&  [13]			&(j)\\ 
\hline
XTEJ1650-500		&27.02		&0.21		&27.479 		 &0.008		&7.69	    &5.65	  &1.65	&50	   &3	 &2.6	   &0.7	 & [19][20]	 	&(o)\\
\hline
XTEJ1752-223		&27.79		&0.14		&28.385	 	 &0.003		&-		 	&9.50	  &1.5	&-	   &-	 &8.0	   &-	& [14]		&(p)(q)\\
\hline
GRS1758-258		&26.86		&0.38  		&27.33	 	&0.05		&442.80		&6.00	  &+3.4-1.45 	&-	   &-	 &8.5	   &-	&[15]		&(r)\\
\hline
MAXIJ1659-152	&27.52		&0.15		&28.309 	  	&0.004		&-			&20.00	  &3		&-	   &-	 &8.0	   &0.0		& [16]			    &(s)\\
\hline
CygX-1  		    &27.51		&0.15		&28.22 	   	&0.12			&134.40     &14.80	  &1		&35	   &5	 &2.1	   &0.4		&[17]    &(b)\\ 
\hline
\end{tabular}

\begin{flushleft}
\footnotesize{\textbf{RefA}:
(1):\cite{Munoz2008}, (2): \cite{Casares1996}, (3):\cite{Wagner1992}, (4): \cite{Shahbaz2003}, (5): \cite{Altamirano2011}, (6): \cite{Rebusco2012}, 	
(7): \cite{Cantrell2010}, (8): \cite{Orosz2003}, (9): \cite{Gelino2003}, (10): \cite{Shaposhnikov2009}	, (11): \cite{Steeghs2013}, (12): \cite{Khargharia2013},  (13): \cite{Orosz2011}, (14):\cite{Shaposhnikov2010}, (15): \cite{Nate2006}, (16): \cite{Shaposhnikov2011}, (17): \cite{Orosz2011b}, (18) \cite{Neustroev2014}, (19): \cite{Orosz2004}, (20): \cite{Ozel2010}. The references of the radio and X-ray luminosities, \textbf{RefB}:
(a)\cite{Corbel2003} (b)\cite{Corbel2013} (c)\cite{Coriat2011} (d)\cite{Corbel2008} (e)\cite{Rushton2010} (f)\cite{Coriat2011b} (g)\cite{Calvelo2010} (h)\cite{Rodriguez2011} (i) \cite{Gallo2006} (j) \cite{Gallo2003}  (k)\cite{Cadolle2007} (l)\cite{Soleri2010} (m) \cite{Coriat2011} (n) \cite{Fender2010} (o)\cite{Corbel2004}  (p)\cite{Brocksopp2013} (q)\cite{Ratti2012} (\cite{Gallo2003}) (s) \cite{Jonker2012}. }
\end{flushleft}


\label{table2.2}
\end{table*}
\end{landscape}


\appendix
\label{lastpage}

\end{document}